\documentclass[a4paper,fleqn,usenatbib]{mnras}

\usepackage[T1]{fontenc}
\usepackage{ae,aecompl}
\usepackage[utf8]{inputenc}

\usepackage{graphicx}	
\usepackage{amsmath}	
\usepackage{amssymb}	
\usepackage{lscape}
\usepackage{gensymb}
\usepackage{import}
\usepackage{lmodern}
\RequirePackage{fix-cm}

\usepackage{graphicx}	
\usepackage{amsmath}	
\usepackage{amssymb}	
\usepackage{grffile}

\newcommand{\rlight}{r_{\rm L}}

\newcommand{\me}{m_{\rm e}}

\newcommand{\rot}{\mathbf{\nabla} \times}
\newcommand{\divg}{\mathbf{\nabla}\cdot}

\title[Radiative pulsar magnetospheres]{Radiative pulsar magnetospheres: aligned rotator}

\author[J\'er\^ome P\'etri]{
J. P\'etri\thanks{E-mail: jerome.petri@astro.unistra.fr}
\\
Universit\'e de Strasbourg, CNRS, Observatoire astronomique de Strasbourg, UMR 7550, F-67000 Strasbourg, France.
}

\date{Accepted XXX. Received YYY; in original form ZZZ}

\pubyear{2019}

\begin{document}
\label{firstpage}
\pagerange{\pageref{firstpage}--\pageref{lastpage}}
\maketitle

\begin{abstract}
Force-free neutron star magnetospheres are nowadays well known and found through numerical simulations. Even extension to general relativity has recently been achieved. However, those solutions are by definition dissipationless, meaning that the star is unable to accelerate particles and let them radiate any photon. Interestingly, the force-free model has no free parameter however it must be superseded by a dissipative mechanism within the plasma. In this paper, we investigate the magnetosphere electrodynamics for particles moving in the radiation reaction regime, using the limit where acceleration is fully balanced by radiation, also called Aristotelian dynamics. An Ohm's law is derived, from which the dissipation rate is controlled by a one parameter family of solutions depending on the pair multiplicity~$\kappa$. The spatial extension of the dissipation zone is found self-consistently from the simulations. We show that the radiative magnetosphere of an aligned rotator tends to the force-free regime whenever the pair multiplicity becomes moderately large, $\kappa \gg 1$. However, for low multiplicity, a substantial fraction of the spindown energy goes into particle acceleration and radiation in addition to the Poynting flux, the latter remaining only dominant for large multiplicities. We show that the work done on the plasma occurs predominantly in the equatorial current sheet right outside the light-cylinder.
\end{abstract}




\section{Introduction}

Neutron stars are well known to emit a broadband electromagnetic spectrum from the radio wavelength \citep{manchester_australia_2005} through optical/UV up to high and very high-energy, in GeV/subTeV \citep{abdo_second_2013}, and even TeV for the Crab pulsar \citep{ansoldi_teraelectronvolt_2016}. However, it is still unclear where precisely in the magnetosphere or the wind these photons are coming from and how they are produced. Ultra-relativistic particles must flow around the neutron star, emitting curvature, synchrotron and/or inverse Compton radiation. Therefore, particle acceleration and its subsequent radiation mechanism cannot be dissociated from the magnetosphere electrodynamics. It is compulsory to self-consistently solve for Maxwell equations and particle dynamics and radiation to attempt to faithfully and confidently reproduce the wealth of multi-wavelength observations.

The zeroth order approximation is the force-free electrodynamics (FFE), no dissipation is allowed and it is assumed that enough particles are produced to efficiently and completely screen the electric field component along the magnetic field. These solutions are now well known for more than a decade thanks to the advent of numerical simulations pioneered by \cite{contopoulos_axisymmetric_1999} for the aligned rotator and later by \cite{spitkovsky_time-dependent_2006} for the oblique rotator. Since then, these results have been confirmed by several other groups, see for instance \cite{timokhin_force-free_2006, komissarov_simulations_2006, petri_pulsar_2012, kalapotharakos_three-dimensional_2009, parfrey_introducing_2012, chen_electrodynamics_2014, tchekhovskoy_three-dimensional_2016} and references therein. Alternative models without current sheets have been found by \cite{lovelace_jets_2006}. Although the plasma is made of electron-positron pairs requiring a two-fluid model, \cite{beskin_particle_2000} showed that for large pair multiplicity, a one-fluid MHD approximation is valid.

Obviously, going beyond this force-free regime is required to properly address the problem of particle acceleration and radiation. Soon after these detailed computations of the force-free magnetospheres, so called resistive magnetospheres have been introduced by modifying the force-free current to take into account a kind of resistivity leading to an Ohm's law showing the plasma reaction to an applied external field. The way to design this law is multifold, no unique prescription has been found \citep{li_resistive_2012, kalapotharakos_toward_2012, gruzinov_strong-field_2008}. In these models, one parameter identified as a kind of resistivity is introduced to control the rate of dissipation making them able to switch from vacuum to a fully force-free magnetosphere. The origin and physical motivation for these resistivities is not clear. \cite{gruzinov_aristotelian_2013} followed another track, trying to compute dissipative magnetospheres by using the radiation reaction limit for an aligned rotator. An Ohm's law is easily derived from radiation reaction. \cite{contopoulos_are_2016}, based on this idea, computed radiative magnetospheres for an oblique rotator. However he employed a simplified prescription for the current, reminiscent of several force-free codes, and not exactly reflecting the true dynamics of radiation reaction (the current flowing along the magnetic field lines was not included). It is expected that gamma-ray light-curve fitting with Fermi/LAT pulsars will help constraining the dissipative regions and emission mechanisms \citep{kalapotharakos_gamma-ray_2014, kalapotharakos_fermi_2017, brambilla_testing_2015}. Dissipation in the equatorial current sheet is the key to our understanding of the pulsar magnetospheres \citep{contopoulos_new_2014}.

This paper computes radiative pulsar magnetospheres for an aligned rotator in the radiation reaction limit by taking into account the full dissipative current: the electric drift component as well as the contributions from the components aligned with respectively the electric and magnetic field. In Sec.~\ref{sec:Modele}, we describe the model of our radiative magnetosphere and the prescription for Ohms law in the radiation reaction regime. Then some examples of field lines are presented in Sec.~\ref{sec:LigneChamp} for FFE and radiative magnetospheres with fixed pair multiplicity. Next, in Sec.~\ref{sec:Luminosite} we compute the spin-down luminosity extracted from these models and compare it with previous works. The importance of dissipation is pointed out in Sec.~\ref{sec:Dissipation}. Conclusions are drawn in Sec.~\ref{sec:Conclusion}.

\section{Magnetospheric model}
\label{sec:Modele}

In the radiative regime, in the same way as in the force-free regime, the plasma inertia and pressure is neglected. The plasma only furnishes the required charge~$\rho_{\rm e}$ and current~$\mathbf{j}$ density to evolve Maxwell equations written in standard MKSA units as
\begin{subequations}
	\begin{align}
	\label{eq:Maxwell1}
	\divg \mathbf B & = 0 \\
	\label{eq:Maxwell2}
	\rot \mathbf E & = - \frac{\partial \mathbf B}{\partial t} \\
	\label{eq:Maxwell3}
	\divg \mathbf E & = \frac{\rho_{\rm e}}{\varepsilon_0} \\
	\label{eq:Maxwell4}
	\rot \mathbf B & = \mu_0 \, \mathbf j + \frac{1}{c^2} \, \frac{\partial \mathbf E}{\partial t}  .
	\end{align}
\end{subequations}
There are no evolution equations for the plasma as in the FFE case. Its charge density~$\rho_{\rm e}$ is deduced from Maxwell-Gauss law. Following the derivation given by \cite{petri_theory_2016}, the radiative current density~$\mathbf{j}$ is expressed solely in terms of the electromagnetic field via
\begin{equation}
\label{eq:courant}
\mathbf j = \rho_{\rm e} \, \frac{\mathbf E \wedge \mathbf B}{E_0^2/c^2 + B^2} + (|\rho_{\rm e}|+2\,\kappa\,n_0\,e) \, \frac{E_0 \, \mathbf E/c^2 + B_0 \, \mathbf B}{E_0^2/c^2 + B^2}
\end{equation}
where $E_0$ and $B_0$ are the strength of the electric and magnetic field deduced from the electromagnetic invariants and satisfying $\mathcal{I}_1 = \bmath E^2 - c^2 \, \bmath B^2 = E_0^2 - c^2 \, B_0^2$ and $\mathcal{I}_2 = c \, \bmath E \cdot \bmath B = E_0 \, B_0$. Explicitly solving for $E_0\geq0$ and $B_0$ we find
\begin{subequations}
	\label{eq:E0B0}
	\begin{align}
	E_0^2 & = \frac{1}{2} \, (\mathcal{I}_1 + \sqrt{\mathcal{I}_1^2 + 4 \, \mathcal{I}_2^2 }) \\
	c\,B_0 & = \textrm{sign} (\mathcal{I}_2) \, \sqrt{E_0^2 - \mathcal{I}_1} .
	\end{align}
\end{subequations}
At large distances from the neutron star, $B_0 \ll B$ and hence, the second term in Eq.~(\ref{eq:courant}) does not play a leading role for not too large multiplicity factors~$\kappa$.
If the force-free condition is satisfied, Eq.~(\ref{eq:E0B0}) reduce to $E_0=0$ and $c\,B_0 = \sqrt{- \mathcal{I}_1}$. $B_0$ is then the magnitude of the magnetic field in the frame where $\mathbf{E}$ vanishes.
$\kappa$ is the pair multiplicity and $n_0>0$ a fiducial particle density number depending on space and time. For concreteness, we set this background particle density to $|\rho_{\rm e}| = n_0\,e$ as done by \cite{gruzinov_pulsar_2013-1} and by \cite{contopoulos_are_2016} although other less restrictive prescriptions are possible at the expense of adding new parameters. In this case the magnitude of the electric current along $\mathbf{E}$ and $\mathbf{B}$ is controlled by the factor $(1+2\,\kappa)$ and can be increased without bounds. Note the important fact that there is no constraint on the magnitude of~$E$ to be less than $c\,B$ nor any constraint on $\mathbf{E} \cdot \mathbf{B}$ any more. However, in magnetically dominated regions where $E<c\,B$ is satisfied, we enforce the force-free condition. This renders the magnetic field lines closing within the light-cylinder inert, not contributing to acceleration or radiation of particles in the dead zone.

In FFE, pairs move in the same direction along $\mathbf{B}$ because there exist no accelerating electric field~$E_\parallel$ to separate them. They show a simple ballistic motion along $\mathbf{B}$. In Aristotelian dynamics, an $E_\parallel$ is allowed, pairs initially moving in the same direction will be separated by reverting the velocity sign for one species depending on its charge. The distance required to revert the velocity remains small compared to~$\rlight$. Indeed it can be estimated by the distance~$s$ onto which the work done by the parallel electric field~$E_\parallel$ equals the kinetic energy $q\,E_\parallel\,s=\gamma\,\me\,c^2$. This leads to $s/\rlight=\gamma/a$ where $a=q\,E_\parallel/m\,c\,\Omega \gg 1$ is the strength parameter of the electromagnetic field. Actually, $E_\parallel$ is simply $E_0$ in the drift frame where $\mathbf{E}$ and $\mathbf{B}$ are parallel. From the simulations shown below we found that $E_0\approx0.1\,E$ in the dissipative region close to the equator even if the drift speed remains sub-relativistic there and therefore $E_0 \approx E_\parallel$. A strong parallel electric field is easily allowed in the radiation reaction limit even close to the light-cylinder. This contrasts with the expectation from FFE where such behaviour is only expected asymptotically at large distances. For typical pulsar parameters we have $\gamma\ll a$ therefore $s\ll \rlight$ meaning that pairs move in opposite direction after travelling a short distance in the same direction. The situation is reminiscent of pair creation in the polar caps. $E_\parallel$ separates both species after a short distance. Radiation reaction regime extends this picture a priori to the full magnetosphere and wind. Numerical simulations effectively decide where exactly the gap physics has to set in. \cite{petri_pulsar_2019} also showed that particles readjust quickly their velocity to conform to Aristotelian dynamics within a small distance much smaller than any macroscopic length-scale supporting this approximation. In other words, the particle velocity at some location is not affected by the electromagnetic field the particle encountered at another previous position.

The procedure to solve for the radiative magnetosphere is the following. Impose a centred dipole at the stellar surface by enforcing continuity of the normal component of $\mathbf{B}$ and continuity of the tangential component of $\mathbf{E}$. Start with a static dipole without electric field outside the star and let it rotate at a speed~$\Omega$ at time $t=0$. Compute the charge density~$\rho_{\rm e}$ to be put in the electric current expression, after computation of the electromagnetic field strengths~$E_0$ and $B_0$. Then solve Maxwell equations to the next time step. Find the new electromagnetic field at the next time step and restart the process from the beginning. The procedure is very similar to FFE simulations, we only add a new parameter~$\kappa$ with a new prescription for the current. We stress that in this radiative model, as in FFE, charge conservation is insured by the computation of the charge density~$\rho_{\rm e}$ from the electric field $\mathbf{E}$ and not the opposite where $\rho_{\rm e}$ would be explicitly evolved through the fluid motion, changing the longitudinal part of $\mathbf{E}$ by a correction algorithm to enforce charge continuity. Meanwhile when reaching the stationary state $\partial_t=0$, the current density eq.~(\ref{eq:courant}) automatically relaxes to the constrain $\divg j=0$ similarly to the FFE case, even if the divergencelessness condition is not satisfied or imposed initially. We also checked a posteriori that this is indeed the case.

We performed several runs with force-free and radiative magnetospheres. The neutron star radius is set to $R/\rlight=0.2$ and the outer boundary of the simulation sphere is located at $5\,\rlight$ where the light-cylinder is defined by $\rlight=c/\Omega$. Moreover, the pair multiplicity factor is chosen in the set~$\kappa=\{0,1,5\}$. An absorbing sponge layer of thickness 10\% of the radial extent is implemented in the outer part in order to avoid small spurious reflections introduced by the outgoing wave boundary conditions of the characteristic compatibility method \citep{canuto_spectral_2007}. Therefore, the results at distances larger than $r\geq4.5\,\rlight$ should be discarded. In the subsequent sections, we discuss the main results about field line structure, spindown losses and work done on the plasma.

\section{Field lines}
\label{sec:LigneChamp}

\begin{figure}
	\centering
	\includegraphics[width=0.35\textwidth]{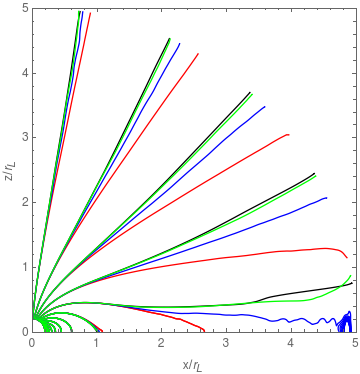} 
	\caption{Magnetic field lines for the force-free magnetosphere in black solid lines, and radiative magnetosphere with $\kappa \in \{0,1,5\}$ in respectively red, blue and green solid lines as shown in the legend.}
	\label{fig:ligne_champ_r0.2_j}
\end{figure}

The geometry of magnetic field lines in the meridional plane is shown in Fig.~\ref{fig:ligne_champ_r0.2_j} for the FFE regime and the radiative regime for several values of the pair multiplicity factor~$\kappa \in \{0,1,5\}$. Inside the light-cylinder, the magnetic field remains insensitive to the current prescription, there is no need for dissipation as $E<c\,B$ within the light-cylinder. At large distances, well outside the light-cylinder, all cases show a split monopole structure but the radiative case tends to close more field lines along the equator. As particle are no more constrained to follow field lines due to acceleration along the electric field~$\mathbf{E}$, we observe diffusion of particle across these lines in adjunction to some dissipation. Indeed, the radiative magnetosphere impacts on largest in the vicinity of the equatorial current sheet. We will show below that work on particles is done essentially in this plane, allowing magnetic field lines to reconnect easily.

Next we diagnose quantitatively the effect of a radiative magnetosphere by computing relevant physical parameters such as the spindown luminosity and the work done on the plasma.

\section{Poynting flux}
\label{sec:Luminosite}

In the ideal limit of a force-free magnetosphere, all the spindown goes into the electromagnetic flux. No energy is carried away by particles, only the Poynting vector propagates radially outwards with constant flux, forming a current sheet in the equatorial plane. Numerically such current sheet is difficult to handle and all schemes have to resort to some dissipation by decreasing artificially the electric field to respect the $E<c\,B$ condition. Unfortunately, this is an uncontrolled process difficult to interpret physically. In the proposed radiative model, dissipation is naturally allowed by assuming particles reaching an exact balance between acceleration and radiation reaction. Motion across field line is then permitted and the two important constraints $\mathbf{E}\cdot\mathbf{B}=0$ and $E<c\,B$ disappear.

The local energy conservation law for the plasma-magnetosphere system, i.e. field and matter, reads
\begin{equation}
\label{eq:ConservationEnergie}
\frac{\partial u}{\partial t} + \divg \mathbf{S} + \mathbf{j} \cdot \mathbf{E} = 0
\end{equation}
where we introduced the electromagnetic energy density by its simplest form
\begin{equation}
\label{eq:DensiteElectromagnetic}
u = \frac{\varepsilon_0 \, E^2}{2} + \frac{B^2}{2\,\mu_0}
\end{equation}
the Poynting flux by
\begin{equation}
\label{eq:FluxPoynting}
 \mathbf{S} = \frac{\mathbf{E} \wedge \mathbf{B}}{\mu_0}
\end{equation}
and the work done on the plasma represented by the last term $\mathbf{j} \cdot \mathbf{E}$. This term vanishes for force-free plasma and in the radiation reaction limit it reduces to
\begin{equation}
\label{eq:jscalaireE}
\mathbf{j} \cdot \mathbf{E} = |\rho_e| \, ( 1 + 2 \, \kappa ) \, c \, E_0 \geq 0 .
\end{equation}
In a stationary state, the electromagnetic energy density~$u$ does not vary. Without dissipation, the Poynting flux across a closed surface is conserved but when energy is deposited into the plasma, it decreases radially outward. Indeed, Eq.~(\ref{eq:ConservationEnergie}) integrated within a sphere~$\Sigma$ of radius~$r$ implies
\begin{equation}
\label{eq:Travail}
L = \iint_\Sigma S_{\rm r} \, d\Sigma = - \iiint_V \mathbf{j} \cdot \mathbf{E} \, dV
\end{equation}
where $S_{\rm r}$ is the radial component of the Poynting flux, $d\Sigma$ a surface element on the sphere and $dV$ a volume element inside the sphere~$\Sigma$.

\begin{figure}
	\centering
	\includegraphics[width=0.45\textwidth]{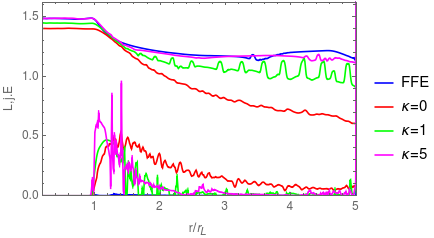} 
	\caption{Radial decrease of the Poynting flux depending on the model. FFE is shown in black, and a radiative magnetosphere in red for $\kappa=0$, in blue for $\kappa=1$, and in green for $\kappa=5$. The associated work is shown in the lower curves.}
	\label{fig:Luminosite}
\end{figure}

The radial evolution of the Poynting flux is shown in the upper curves of Fig.~\ref{fig:Luminosite}. The luminosity is normalized with respect to the vacuum orthogonal rotator $\ell = L/L_{\rm vac}$. The black solid line correspond to the FFE magnetosphere with a spindown equal to $\ell \approx1.469$. The work done on the plasma is null, however, because the magnetically dominant case must be satisfied, we artificially decrease the electric field strength. This happens mostly outside the light-cylinder and is visible as a sensitive decrease of about 20\% in the luminosity already in FFE.

For radiative magnetosphere, dissipation already occurs inside the light cylinder as seen by inspecting the red, blue and green solid curves. Note that this effect is physical, not numerical. For $\kappa=0$, red curve, we observed a dissipation rate of about 5\% inside the light-cylinder with a Poynting flux $\ell\approx1.399$ but up to 50\% outside. This energy dissipation into particle is similar to the results found by PIC simulations by \cite{belyaev_dissipation_2015} when injections of pairs is weak and localized inside the magnetosphere. Surprisingly, even for high pair multiplicity which should converge to the force-free solution, a significant fraction of the Poynting flux, about 30\% is injected into particles \citep{cerutti_particle_2015}. In our simulations, electromagnetic work done on the plasma is most efficient in electrically dominant regions, i.e. in the equatorial current sheet. When pairs are added into the magnetosphere, the dynamics tends to the FFE case. For $\kappa=1$ dissipation is almost reduce by a factor~2, blue curve $\ell \approx1.431$. For $\kappa=5$, green curve, dissipation becomes very weak and the magnetosphere resembles closely to the FFE model with the same Poynting flux at the light-cylinder of $\ell \approx  1.459$.

\begin{figure}
	\centering
	\includegraphics[width=0.45\textwidth]{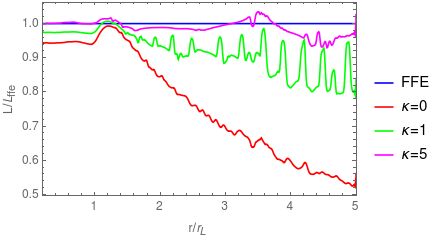} 
	\caption{Relative Poynting flux normalized to the FFE spindown and depending on the pair multiplicity~$\kappa$. The radiative magnetosphere for $\kappa=0$ is shown red, for $\kappa=1$ in blue and for $\kappa=5$ in green.}
	\label{fig:LuminositeRelative}
\end{figure}

We think the relative luminosity compared to the FFE case computed by $L/L_{\rm ffe}$ as shown in Fig.~\ref{fig:LuminositeRelative} better reflects the true physical dissipation introduced by radiation reaction. Indeed, the FFE magnetosphere should not dissipate but due to the numerical algorithm, we always find a significant dissipation outside the light cylinder that cannot be reduced by increasing the resolution. The radiative losses are seen to be significant outside the light-cylinder.

\section{Dissipation}
\label{sec:Dissipation}

Let us carefully examine the dissipation process, converting electromagnetic energy into particle acceleration and radiation. The decrease in the Poynting flux~$L$ with distance is imputed to the work done on the plasma. It is very informative to check where exactly in the magnetosphere or wind the electromagnetic energy is transferred to the particles. The local work done on the plasma within a spherical shell, expressed as
\begin{equation}
\label{eq:DeriveeTravail}
\iint_\Sigma \mathbf{j} \cdot \mathbf{E} \, d\Sigma = - \frac{dL}{dr}
\end{equation}
is shown in the lower curves of Fig.~\ref{fig:Luminosite}. It shows how fast dissipation occurs when leaving the magnetosphere. In the FFE case, black line, it should be exactly zero, but because of numerical filtering and artificial decrease of $E$, a small residual work is seen right at the light-cylinder. For the most dissipative case, $\kappa=0$ in red line, the work done start at the light-cylinder, reaching a maximum around $1.5\rlight$ and then decreases slowly. For $\kappa=1$, blue line, dissipation also start at the light-cylinder, but with a sharp increase and a faster decrease at larger distances becoming very weak at the outer boundary. For $\kappa=5$, green line, dissipation is restricted to the range $[1,1.5]\rlight$, quickly diminishing to negligible values at $r\geqslant2\rlight$. For large multiplicities, we anticipate that dissipation happens only right at the light-cylinder, with a constant Poynting flux at large distances, but different from the Poynting flux inside the light-cylinder. 

\begin{figure}
	\centering
	\includegraphics[width=0.9\linewidth]{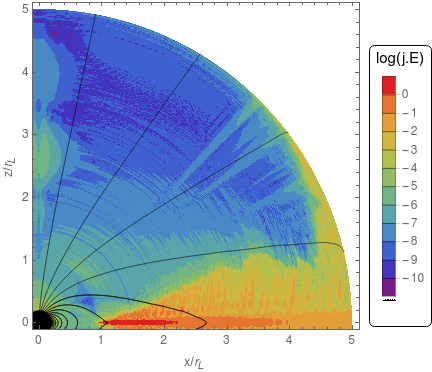}
	\caption{Work done on the plasma for $\kappa=0$ as given by Eq.~(\ref{eq:jscalaireE}).
	}
	\label{fig:travail}
\end{figure}

\begin{figure}
	\centering
	\includegraphics[width=0.9\linewidth]{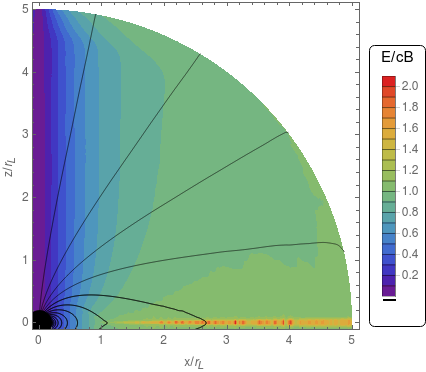}
	\caption{Electric to magnetic field strength ratio~$E/cB$ for $\kappa=0$.}
	\label{fig:EsB}
\end{figure}

In order to better localize the radiative regions, we plot a map of the work done locally on the plasma by evaluating the power in Eq.~(\ref{eq:jscalaireE}). The most interesting case is shown in Fig.~\ref{fig:travail} for $\kappa=0$. As expected, Poynting flux flows into the plasma mainly along the equatorial current sheet, right at the light-cylinder. After several $\rlight$, the power sharply decreases by two orders of magnitude at the outer boundary. The maximum thickness of this dissipative region is about $0.2\,\rlight$. No relevant work is done on the plasma outside the equatorial plane.
Finally, we plot the electric to magnetic field strength ratio~$E/cB$ for $\kappa=0$ in Fig.~\ref{fig:EsB}. Within the light-cylinder, the situation is identical to a FFE magnetosphere because naturally $E<c\,B$. However the electric field dominates in the current sheet being 1.6~times larger than $c\,B$. These are the places where particle acceleration and radiation happen.

\section{Conclusions}
\label{sec:Conclusion}

From an observational point of view, neutron star magnetospheres must be dissipative in order to accelerate particles and radiate photons. We showed that radiation reaction in the ultra-relativistic regime introduces a kind of resistivity self consistently with a map of acceleration and radiation zones. The magnetosphere adjusts itself to a new equilibrium state where most of the dissipation occurs in the equatorial current sheet outside the light-cylinder. We found that the efficiency of dissipation is related to the pair multiplicity factor~$\kappa$. When the pair supply is high enough, the radiative magnetosphere tends again to the force-free solution.

However, imposing a charge density according to a constant pair multiplicity factor and to the local electric field via Maxwell-Gauss law is too restrictive. In a next step, we plan to add explicitly a source of electron-positron pairs to be deposited along the polar caps and/or within the whole magnetosphere. The particle density number then satisfies a conservation law for each species to be solved in addition to Maxwell equation.

Moreover, going to full 3D radiative magnetosphere is mandatory in order to predict phase resolved spectra and light-curves, facilitating the comparison with the wealth of multi-wavelength observations of pulsars. This should help to constrain the dynamics of the equatorial current sheet, also called striped wind in the literature and to get physical insight into the electrodynamics of radiating pulsar magnetospheres.

\section*{Acknowledgements}

We are grateful to the referee for helpful comments and suggestions. This work has been supported by the CEFIPRA grant IFC/F5904-B/2018. We acknowledge the High Performance Computing center of the University of Strasbourg for supporting this work by providing scientific support and access to computing resources. 








%


\bsp	
\label{lastpage}
\end{document}